\documentclass[twocolumn,showpacs,showkeys,preprintnumbers,amsmath,amssymb,prd]{revtex4}

\usepackage{graphicx}
\usepackage{dcolumn}
\usepackage{bm}
\tighten

\begin{document}

\title{On representations of the rotation group and magnetic
monopoles}

\author{Alexander I. Nesterov}
   \email{nesterov@cencar.udg.mx}
\author{F. Aceves de la Cruz}%
 \email{fermin@udgphys.intranets.com}
\affiliation{Departamento de F{\'\i}sica, CUCEI, Universidad de
Guadalajara, Av. Revoluci\'on 1500, Guadalajara, CP 44420, Jalisco,
M\'exico}

\date{\today}
\begin{abstract}
Recently (Phys. Lett. {\bf A302} (2002) 253, hep-th/0208210; hep-th/0403146) employing bounded
infinite-dimensional representations of the rotation group we have
argued that one can obtain the consistent monopole theory with
generalized Dirac quantization condition, $2\kappa\mu \in \mathbb
Z$, where $\kappa$ is the weight of the Dirac string. Here we
extend this proof to the unbounded infinite-dimensional
representations.
\end{abstract}



\maketitle


\section{Introduction}

The Dirac quantization relation \cite{Dir} between an electric
charge $e$ and magnetic charge $q$,
\begin{equation}\label{D1}
2\mu =n, \; n \in \mathbb Z,
\end{equation}
where $\mu =eq$, and we set $\hbar = c=1$, has been obtained from
various approaches based on quantum mechanics and quantum field
theory \cite{Dir,T,Gol1,Gold,Gol2,Sch,H,LWP,Zw_1,Wu1,Wu2}.  One of
the widely accepted proofs of the Dirac selection rule is based on
group representation theory and consists in the following: In the
presence of magnetic monopole the operator of the total angular
momentum
\begin{equation}
{\mathbf J} = {\mathbf r} \times \left({-i\mathbf \nabla} -
e{\mathbf A}\right) - \mu\frac{\mathbf r}{r},
\end{equation}
has the same properties as a standard angular momentum and for any
value of $\mu$ obeys the usual commutation relations
\begin{eqnarray}
[J_i, J_j] = i\epsilon_{ijk}J_k. \label{eq5}
\end{eqnarray}
The requirement that $J_i$ generate a finite-dimensional
representation of the rotation group yields $2\mu$ being integer
and only values $2\mu = 0, \pm1,\pm2, \dots$ are allowed (for
details see, for example, \cite{Gol1,Gol2,H,LWP,Zw_1}).

Actually the charge quantization does not follow from the
quantum-mechanical consideration and rotation invariance alone.
Any treatment uses some additional assumptions that may be not
physically inevitable.

Recently we have exploited this problem  employing bounded
infinite-dimensional representations of the rotation group and
nonassociative gauge transformations. We argued that one can relax
Dirac's condition and obtain the consistent monopole theory with
the generalized quantization condition, $2\kappa\mu \in \mathbb
Z$, $\kappa$ being the weight of the Dirac string \cite{NA,NA1}.
In our Letter we extend this proof to the unbounded
infinite-dimensional representations of the rotation group.

\section{Magnetic monopole preliminaries}

As well-known any  vector potential $\mathbf A$ being compatible
with a magnetic field ${\mathbf B}= q {\mathbf r}/r^3$ of Dirac
monopole must be singular on the string (the so-called {\em Dirac
string}, further it will be denoted as $S_{\mathbf n}$), and one
can write
\[
{\mathbf B}={\rm rot}{\mathbf A}_{\mathbf n} + {\mathbf
h}_{\mathbf n}
\]
where ${\mathbf h}_{\mathbf n}$ is the magnetic field of the Dirac
string  given by
\begin{equation}
{\mathbf h}_{\mathbf n}  = 4\pi q{\mathbf n}\int _{0}^\infty
\delta^3(\mathbf r - \mathbf n \tau) d \tau.
\end{equation}
The unit vector $\mathbf n$ determines the direction of a string
$S_{\mathbf n}$ passing from the origin of coordinates to
$\infty$.

For instance, Dirac's original vector potential reads
\begin{equation}
{\mathbf A}_{\mathbf n}= q\frac{{\mathbf r}\times {\mathbf n}}
{r(r - {\mathbf n} \cdot{\mathbf r})} \label{d_str},
\end{equation}
and the Schwinger's choice is
\begin{equation}
{\mathbf A^{SW}}= \frac{1}{2}\bigl({\mathbf A}_{\mathbf n}+
{\mathbf A}_{-\mathbf n} \bigr), \label{sw}
\end{equation}
the string being propagated from $-\infty$ to $\infty$ \cite{Sch}.
Both vector potentials yield the same magnetic monopole field,
however the quantization is different. The Dirac condition is
$2\mu=p$, while the Schwinger one is $\mu=p, \; p\in \mathbb Z$.

These two strings belong to a family $\{{\mathbf
S}^{\kappa}_{\mathbf n} \}$ of  {\it weighted strings}, $\kappa$
being the weight of the semi-infinite Dirac string \cite{NA,NA1}.
The respective vector potential is defined as
\begin{align}
{\mathbf A}^\kappa_{\mathbf n} = \kappa {\mathbf A}_{\mathbf n} +
(1 - \kappa){\mathbf A}_{-\mathbf n}, \label{A_1}
\end{align}
and the  magnetic field  of the string ${\mathbf
S}^{\kappa}_{\mathbf n}$ is
\begin{align}
&{\mathbf h}^{\kappa}_{\mathbf n}=  \kappa{\mathbf h}_{\mathbf n}
+ (1-\kappa){\mathbf h}_{-\mathbf n} \label{str}
\end{align}
Since ${\mathbf A}^\kappa_{-\mathbf n}={\mathbf
A}^{1-\kappa}_{\mathbf n}$, we obtain the following equivalence
relation: ${S}^{\kappa}_{-\mathbf n} \simeq{S}^{1-\kappa}_{\mathbf
n} $.

Two strings $S^\kappa_{\mathbf n}$ and $S^\kappa_{\mathbf n'}$ are
related by the gauge transformation
\begin{eqnarray}
A^{\kappa'}_{\mathbf n'}= A^\kappa_{\mathbf n}+ d\chi.
\label{ag2a}
\end{eqnarray}
and vice versa. An arbitrary transformation of the strings
$S^\kappa_{\mathbf n} \rightarrow S^{\kappa'}_{\mathbf n'}$ can be
realized as combination of $S^\kappa_{\mathbf n} \rightarrow
S^{\kappa}_{\mathbf n'}$ and $S^\kappa_{\mathbf n} \rightarrow
S^{\kappa'}_{\mathbf n}$, where the first transformation is
rotation, and the second one results in changing of the weight
string $\kappa \rightarrow \kappa'$ without changing its
orientation.

Let denote by $\mathbf n'= g\mathbf n , g\in\rm SO(3)$, the left
action of the rotation group induced by $S^\kappa_{\mathbf n}
\rightarrow S^\kappa_{\mathbf n'}$. From rotational symmetry of
the theory it follows this gauge transformation $S^\kappa_{\mathbf
n} \rightarrow S^{\kappa}_{\mathbf n'}$ can be undone by rotation
$\mathbf r \rightarrow  \mathbf r g$ as follows
\begin{align}
&A^{\kappa}_{\mathbf n'}(\mathbf r)= A^{\kappa}_{\mathbf
n}(\mathbf r')= A^\kappa_{\mathbf n}(\mathbf r)+ d\alpha(({\mathbf
r}; g)) ,
\label{g_0}\\
&\alpha(\mathbf r;g)= e \int_{\mathbf r}^{\mathbf r'} \mathbf
A^\kappa_{\mathbf n}(\boldsymbol \xi) \cdot d \boldsymbol \xi,
\quad \mathbf r' =  \mathbf r g \label{g_1c}
\end{align}
where the integration is performed along the geodesic
$\widehat{\mathbf r \,\mathbf r'}\subset S^2$.

The transformation of the string $S^\kappa_{\mathbf n} \rightarrow
S^{\kappa'}_{\mathbf n}$ is given by
\begin{eqnarray}
&&A^{\kappa'}_{\mathbf n} = A^{\kappa}_{\mathbf n} -
d\chi_{\mathbf n}, \label{A_02a}\\
 &&d\chi_{\mathbf n} = 2q{(\kappa' - \kappa)}\frac{(\mathbf r \times
\mathbf n)\cdot d\mathbf r}{r^2- (\mathbf n \cdot \mathbf r)^2},
\label{A_02}
\end{eqnarray}
where $\chi_{\mathbf n}$ is polar angle in the plane orthogonal to
${\mathbf n}$. This type of gauge transformations being singular
one can be undone by  combination of the inversion $\mathbf
r\rightarrow -\mathbf r$ and  $\mu\rightarrow -\mu$. In
particular, if $\kappa' = 1-\kappa$ we obtain the mirror string:
$S^\kappa_{\mathbf n} \rightarrow S^{\kappa}_{-\mathbf n}\simeq
S^{1-\kappa}_{\mathbf n} $.

\section{Representations of the rotation group and Dirac's quantization
condition}

Let $\psi^\ell_\nu$ be an eigenvector of the operators $J_3$ and
$J^2$:
\begin{equation}\label{J1a}
J_3\psi^\lambda_\nu=\nu\psi^\ell_\nu, \quad J^2\psi^\ell_\nu
=\ell(\ell +1)\psi^\ell_\nu,
\end{equation}
$\nu,\;\ell$ being real numbers. Involving the operators $J_{\pm}=
J_1 \pm J_2$ it is easy to show that the spectrum of $J_3$ has the
form $\nu =\nu_0 +n$, where $n=0,\pm 1,\pm 2,\dots$ .

Each irreducible representation is characterized by an eigenvalue
of Casimir operator and the spectrum of the operator $J_3$. There
are four distinct classes of representations \cite{San1,San2,Wyb}:
\begin{itemize}
\item {\em Representations unbounded from above and below}, in this
case neither $\ell + \nu_0$ nor $\ell - \nu_0$ can be integers.

\item {\em Representations bounded below}, with
$\ell + \nu_0$ being an integer, and $\ell - \nu_0$ not equal to
an integer.

\item {\em Representations bounded above,} with $\ell - \nu_0$ being an
integer, and $\ell + \nu_0$ not equal to an integer.

\item {\em Representations bounded from above and below,} with $\ell -
\nu_0$ and $\ell + \nu_{0}$ both being integers, that yields $\ell
= k/2, \quad k\in \mathbb Z_{+}$.

\end{itemize}
The nonequivalent representations in each of the series of
irreducible representations are denoted respectively by
$D(\ell,\nu_0)$, $D^{+}(\ell)$, $D^{-}(\ell)$ and $D(k/2)$. The
representations $D(\ell,\nu_0)$, $D^{+}(\ell)$ and $D^{-}(\ell)$
are infinite dimensional; $D(k/2)$ is (k+1)-dimensional
representation. The representations $D^{\pm}(\ell)$ and
$D(\ell,\nu_0)$ are discussed in detail in
\cite{AG,BL,S,San1,San2}.

In fact the representations $D{(\ell},\nu) $ and $D({-\ell
-1},\nu) $, yielding the same value  $Q= \ell(\ell +1)$ of the
Casimir operator, are equivalent and the inequivalent
representations may be labeled as $D({Q},\nu) $ \cite{Wyb}. If
there exists the number $p_0\in \mathbb Z$ such that $\nu + p_0=
\ell$, we have $J_{+}|\ell,\ell\rangle = 0$ and the representation
becomes bounded above. In the similar manner if for a number
$p_1\in \mathbb Z$ one has $\nu + p_1=-\ell $, then
$J_{-}|\ell,-\ell\rangle = 0$ and the representation reduces to
the bounded below. Finally, finite-dimensional unitary
representation arises when there exist possibility of finding
$J_{+}|\ell,\ell\rangle = 0$ and $J_{-}|\ell,-\ell\rangle = 0$. It
is easy to see that in this case $2\ell, \;2m$ and $2 \nu$ all
must be integers.

In what  follows we will discuss the Dirac monopole problem within
the framework of the representation theory outlined above.

Taking into account the spherical symmetry of the system, the
vector potential can be taken as \cite{Wu1,Wu2}
\begin{eqnarray}
{A_N} =  q(1-\cos{\theta})d\varphi, \quad {A_S} =
-q(1+\cos{\theta})d\varphi \label{eq1b}
\end{eqnarray}
where $(r,\theta,\varphi)$ are the spherical coordinates, and
while ${A_{N}}$ has singularity on the south pole of the sphere,
${ A_{S}}$ on the north one. In the overlap of the neghborhoods
covering the sphere $S^2$ the potentials ${A_N}$ and ${A_S}$  are
related by the following gauge transformation:
\[
A_S = A_N - 2qd\varphi.
\]
This is the particular case of transformation given by Eq.
(\ref{A_02a}), when $\kappa = 0$ and $\kappa' =1$.

We start by choosing the vector potential as
$$A= q(1-\cos \theta)d\varphi.$$
Then for the operators $J_i$'s we have
\begin{eqnarray}
&&J_{\pm}= e^{\pm
i\varphi}\bigg(\pm\frac{\partial}{\partial\theta} +i\cot\theta
\frac{\partial}{\partial\varphi} -
\frac{\mu\sin\theta}{1+\cos\theta} \bigg),\\
&&J_3=-i\frac{\partial}{\partial\varphi} - \mu,\\
&&{\mathbf J^2} =-\frac{1}{\sin{\theta}}
\frac{\partial~}{\partial\theta}\left(\sin{\theta}
\frac{\partial~}{\partial\theta}\right) -
\frac{1}{\sin^2{\theta}}\frac{\partial^2~}{\partial\varphi^2} +
\nonumber\\
&&+\frac{2i\mu}{1 +\cos{\theta}}\frac{\partial~}{\partial\varphi}
+\mu^2\frac{1 - \cos{\theta}}{1 + \cos{\theta}} +\mu^2
.\label{eq7}
\end{eqnarray}
Substituting the wave function $\Psi= R(r)Y(\theta,\varphi)$ into
Schr\"odinger's equation
\begin{equation}
\hat H\Psi = E \Psi, \label{eq01}
\end{equation}
we obtain for the angular part the following equation:
\begin{eqnarray}
&&{\mathbf J^2}Y(\theta,\varphi) = \ell({\ell} + 1)Y(\theta,
\varphi). \label{eq7a}
\end{eqnarray}
Starting with $J_3 Y = m Y$ and assuming
\begin{eqnarray}
Y =e^{i(m+\mu)\varphi}z^{(m+\mu)/2}(1-z)^{(m-\mu)/2}F(z),
\label{eq8}
\end{eqnarray}
 where $z=(1-\cos\theta)/2$, we obtain the resultant equation in the
standard form of the hypergeometric equation,
\begin{align}
&z(1-z)\frac{d^2F}{dz^2} +\bigl(c-(a+b+1)z\bigr)\frac{dF}{dz}-abF=0, \\
&a = m - \ell, \; b = m + \ell + 1, \; c = m + \mu + 1. \nonumber
\label{hyp1}
\end{align}

The hypergeometric function $F(a,b;c;z)$ reduces to a polynomial
of degree $n$  in $z$ when $a$ or $b$ is equal to $-n, \;(n =
0,1,2, \dots)$ \cite{Abr,And}, and the respective  solution of Eq.(\ref{eq7a}) is
of the form
\begin{equation}\label{pol_2}
Y_n^{(\delta,\gamma)}(u) =C_n\,(1-u)^{\delta/2}(1+u)^{\gamma/2}
P_n^{(\delta,\gamma)}(u),
\end{equation}
$P_n^{(\delta,\gamma)}(u)$ being the Jacobi polynomials,  $u =
\cos\theta$, and the normalization constant $C$ is given by
\begin{equation*}
C_n=\Bigg(\bigg|\frac{2 \pi \, 2^{\delta +\gamma +1}\Gamma(n
+\delta +1) \Gamma(n +\gamma +1)}{\Gamma(n+1)\Gamma(n +\delta
+\gamma+1)}\bigg| \Bigg)^{-1/2}
\end{equation*}
The functions $Y_n^{(\delta,\gamma)}(u)$ form the basis of the
representation bounded above or below. This case has been studied
in detail in \cite{NA,NA1}.

If both of $a$ and $b$ are negative integers, that is $m+
\ell=-p,\; m+\ell =-k, \; p,k \in {\mathbb Z}_+$, then the
representation becomes finite dimensional. It is easy to check
that in this case $m + \mu$ and $m -\mu$ must be integers, that
yields the Dirac quantization condition $2\mu \in \mathbb Z$.

In the rest of the paper we will discuss the representation
$D(\ell,\mu)$ unbounded above and below. We are looking  for the
solutions of the Eq. (\ref{eq7a}) such that  being regular at the
point $z=0$,  in general, can have singularity at $z=1$, where the
Dirac string crosses the sphere. As a result we obtain the
following restrictions on the spectrum of the operator $J_3$:
\begin{equation}\label{Spec1} m + \mu =n, \; n=0,\pm 1,\pm 2,
\dots.
\end{equation}
The respective solution is given by
\begin{align}
&Y^{(\mu,n)}_\ell = C(\ell,\mu,n)e^{i
n\varphi}z^{n/2}(1-z)^{n/2-\mu}F(a,b,c;z),\label{B1}\\
&a = n - \mu - \ell, \; b =n - \mu  + \ell + 1, \; c =1+n,
\nonumber
\end{align}
where $ C(\ell,\mu,n)$  is a suitable normalization constant (for
the details of the normalization procedure see
\cite{NA1,San1,San2}).

Consider now the other choice of the vector potential
$$A= -q(1+\cos \theta)d\varphi,$$
which corresponds to the Dirac string crossing the sphere at north
pole ($z=0$). In this case  the solution
$\tilde{Y}^{(\mu,n)}_\ell$ of the equation (\ref{eq7a}) being
regular at the point $z=1$ takes the same form as in Eq.
(\ref{B1})
\begin{align}
&\tilde{Y}^{(\mu,n)}_\ell = C(\ell,\mu,n)e^{i
n\varphi}z^{n/2+\mu}(1-z)^{n/2}F(a,b,c;1-z),\label{B2}\\
&a = n + \mu - \ell, \; b =n + \mu  + \ell + 1, \; c =1+n.
\nonumber
\end{align}
The spectrum of the operator $J_3$ being diferent from
(\ref{Spec1}) is found to be
\begin{equation}\label{Spec2}
m - \mu =n, \; n=0,\pm 1,\pm 2, \dots.
\end{equation}
Notice that the functions $\tilde{Y}^{(\mu,n)}_\ell$ can be
obtained from ${Y}^{(\mu,n)}_\ell$ by the change of $z \mapsto
(1-z) $ and $\mu \mapsto -\mu$, that is agree with the gauge
transformation  $$A_S = A_N -2qd\varphi$$ (see also
Eqs.(\ref{A_02a}),(\ref{A_02})) .

The set of the functions
$\left\{\tilde{Y}^{(\mu,n)}_{\ell},Y^{(\mu,n)}_{\ell}\right\}$
form the complete bi-orthonormal canonical basis of the
representation $D(\ell,\mu)$ in the indefinite-metric Hilbert
space with the indefinite metric given by \cite{HS}
\begin{equation}\label{eq8a}
 \eta_{m m'} = (-1)^{\sigma(m)}\delta_{m m'},
\end{equation}
where
\begin{eqnarray*}
 (-1)^{\sigma(m)}= {\rm sgn}\big(\Gamma(\ell-m
+1)\Gamma(\ell+m+1)\big),
\end{eqnarray*}
${\rm sgn}(x)$ being the signum function. One can see that the
spectrum of the operator $J_3$ is unbounded, double-degenerate and
discrete.

The general case of an arbitrary weighted string
$S^\kappa_{\mathbf n}$ can be considered in the following way: For
$m\pm\mu =n$ the weighted solutions of the Schr\"odinger equation
are given by
\begin{align}
&Y^{(\mu,n)}_{\kappa,\ell} = {\rm
e}^{-2i\kappa\mu\varphi}Y^{(\mu,n)}_\ell, \quad m=n-\mu
\label{W1}\\
&\tilde{Y}^{(\mu,n)}_{\kappa,\ell} = {\rm
e}^{-2i\kappa\mu\varphi}\tilde{Y}^{(\mu,n)}_\ell, \quad m=n+\mu.
\end{align}
Since a Dirac string may be rotated by gauge transformation the
widely accepted point of view is that the string is unobservable.
Thus, to avoid the appearence of an Aharonov-Bohm effect produced
by a Dirac string, one has to impose the generalized Dirac
quantiztion condition $2\kappa\mu \in \mathbb Z$. In particular
cases $\kappa =1$ and $\kappa=1/2$ it yields the Dirac and
Schwinger selectional rules respectively.

\section{Concluding remarks}

We have argued, by applying infinite-dimensional representations
of the rotation group, that the Dirac quantization condition can
be relaxed and changed by $2\kappa\mu \in \mathbb Z$, where
$\kappa$ is the weight of the Dirac string. This selectional rule
arises as natural condition of being consistent with an algebra of
observables and ensures the absence of an Aharonov-Bohm effect
produced by Dirac string. Moreover, since there is no any
restriction on the parameter $\kappa$, an arbitrary magnetic
charge is allowed.

It follows from our description that the spectrum of the operator
$J_3$ is double-degenerate, discrete and unbounded, $m = n\pm\mu$.
The physical interpretation of this result is not clear yet. We
believe that it can be explained treating the charge-monopole
system as a free anyon with translational and spin degrees of
freedom \cite{NA2}.

\section*{Acknowledgments}

One of the authors, F.A., thanks Center for Theoretical Physics of
the Massachusetts Institute of Technology where the part of this
work  has been done, for the warm hospitality. This work was
supported by UdeG, Grant No. 5025.

\end{document}